\documentclass[traditabstract]{aa} 
\usepackage{graphicx}
\usepackage{txfonts}
\usepackage{natbib}
\bibpunct{(}{)}{;}{a}{}{,}

\begin{document}

   \title{Low albedos of hot to ultra-hot Jupiters in the optical to near-infrared transition regime
\thanks{Based on data obtained with the STELLA robotic telescopes in Tenerife, an AIP facility jointly operated by AIP and IAC.}}

   \titlerunning{Low albedos for five hot Jupiters}

   \author{M.~Mallonn\inst{1}, J. K\"{o}hler\inst{1}, X.~Alexoudi\inst{1}, C.~von Essen\inst{2}, T.~Granzer\inst{1}, K. Poppenhaeger\inst{1}, K.\,G.~Strassmeier\inst{1}}
   \authorrunning{M. Mallonn et al.}

\institute{Leibniz-Institut f\"{u}r Astrophysik Potsdam, An der Sternwarte 16, D-14482 Potsdam, Germany \\ 
  \email{mmallonn@aip.de}
\and 
Stellar Astrophysics Centre, Department of Physics and Astronomy, Aarhus University, Ny Munkegade 120, DK-8000 Aarhus C, Denmark 
}

   \date{Received --; accepted --}

  \abstract{The depth of a secondary eclipse contains information of both the thermally emitted light component of a hot Jupiter and the reflected light component. If the dayside atmosphere of the planet is assumed to be isothermal, it is possible to disentangle both. In this work, we analyzed 11 eclipse light curves of the hot Jupiter HAT-P-32\,b obtained at 0.89~$\mu$m in the z' band. We obtained a null detection for the eclipse depth with state-of-the-art precision, $-0.01\pm0.10$~ppt. We confirm previous studies showing that a non-inverted atmosphere model is in disagreement to the measured emission spectrum of HAT-P-32\,b. We derive an upper limit on the reflected light component, and thus, on the planetary geometric albedo $A_g$. The 97.5\% confidence upper limit is $A_g < 0.2$. This is the first albedo constraint for HAT-P-32\,b, and the first z' band albedo value for any exoplanet. This finding disfavors the influence of large-sized silicate condensates on the planetary day side. We inferred z' band geometric albedo limits from published eclipse measurements also for the ultra-hot Jupiters \mbox{WASP-12\,b}, \mbox{WASP-19\,b}, \mbox{WASP-103\,b}, and WASP-121\,b, applying the same method. These values consistently point to a low reflectivity in the optical to near-infrared transition regime for hot to ultra-hot Jupiters.}

   \keywords{methods: observational --
techniques: photometric -- 
                planets and satellites: fundamental parameters
               }

   \maketitle
%

\section{Introduction}

During the event of a secondary eclipse, an extrasolar planet disappears behind its host star as seen from Earth. This event offers the possibility to differentiate the light originating from the planet from the light of the host star. If observed photometrically, a time series reveals a slight dimming in the light curve during the eclipse. 
The amplitude of the flux dimming is a measure of the flux contribution of the planet to the combined star-planet flux. At optical wavelengths, the planetary light is mostly reflected light of the host star \citep{Winn2008b,Evans2013}. At near-infrared (NIR) wavelengths, however, the thermal emission is expected to dominate the reflected light dependent on the planetary temperature \citep{LopezMorales2007}. In this case, the depth of a secondary eclipse light curve provides information on the brightness temperature of the day side of the planet \citep[see review by][]{Alonso2018}. 
Because this quantity depends on the employed observing bandpass, an emission spectrum of the planet can be built when the secondary eclipse is observed at multiple wavelengths. Such spectrum exhibits fingerprints of the elemental composition and can also reveal information on the planetary temperature-pressure profile \citep{Grillmair2007,Knutson2008,Evans2017,Arcangeli2018,Nikolov2018}.

For close-in gas giants, the so-called hot Jupiters, \cite{LopezMorales2007} described the reflected light component of the secondary eclipse depth to be negligible redward of about 0.8~$\mu$m compared to the thermal emission. Consequently, the reflected light and geometric albedos, i.e., the fraction of light reflected toward the illuminating star, have been mostly determined at optical wavelengths shortward of about 0.7~$\mu$m \citep{Rowe2008,Evans2013,Demory2013,Angerhausen2015,Bell2017}. Very few attempts have been made at longer wavelengths to disentangle reflected from thermal light. One example is the work of \cite{Keating2017}, who showed that secondary eclipse observations in the thermal wavelength domain can reveal very useful constraints on the geometric albedo. 

If available, wavelength information on the geometric albedo can inform about the presence and composition of clouds \citep{Marley1999,Sudarsky2000}. In the NIR, it was possible to deduce albedo information from thermal phase variations and considerations on the energy budget of close-in gas giants \citep{Schwartz2015,Wong2015,Schwartz2017}. These data have shown an apparent disagreement of their rather high Bond albedos (the fraction of stellar energy that is reflected, integrated over phase angle and wavelength) \mbox{$A_B \approx 0.3-0.4$} to the generally low geometric albedos \mbox{$A_g \lesssim 0.2$} at optical wavelengths \citep{Rowe2008, Heng2013}. This offset might be explained by the presence of clouds that reflect NIR radiation, in addition to the existence of optical absorbers \citep{Schwartz2015}.

In this work, we aim to constrain for the first time the geometric albedo in the z' band at 900~nm. This wavelength is intermediate to the existent optical geometric and NIR Bond albedo constraints. In the proposed scenario of \cite{Schwartz2015}, we expect an increase in reflectivity compared to the optical because the opacity of the potential optical absorber TiO and VO is decreased. For this purpose, we analyze new z' band observations of HAT-P-32\,b in this work, and reanalyze existent z' band observations of four other ultra-hot Jupiters, aiming to deduce information on their reflected light component. These other planets are WASP-12\,b, WASP-19\,b, WASP-103\,b, and WASP-121\,b.

Hot Jupiter HAT-P-32\,b is a close-in gas giant of 1750~K equilibrium temperature \citep{Hartman2011}. Its atmosphere at the terminator region, dividing the day side from night side of the planet, was investigated by multiple studies. In this region, the atmosphere is opaque at optical wavelengths because of a cloud layer at high altitudes \citep{Gibson2013,MallonnStrass,Nortmann2016} with indications of an additional haze layer causing scattering \citep{Mallonn2017, Tregloan2018}. At NIR wavelengths, a water feature is measured, whose low amplitude might be caused by a cloud layer muting the spectral signature \citep{Damiano2017,Fisher2018}.

\cite{Zhao2014} and \cite{Nikolov2018} provided an emission spectrum of the day side of the planet by the photometric observation of the secondary eclipse at wavelengths from 1.1 to 4.5~$\mu$m. Because the planetary emission spectroscopy probes deep into the atmosphere, the resulting spectrum is sensitive to the temperature-pressure (P-T) profile of the atmosphere. Both studies consistently found indications for a profile that is either isothermal (resembling a blackbody spectrum) or exhibits a weak temperature inversion, i.e., the temperature rises with higher altitudes. The measured spectrum ruled out strong molecular absorption features of, for example, water. \cite{Zhao2014} used their data to determine a circular orbit for the planet. 

A nearby stellar object to HAT-P-32 was found by \cite{Adams2013} and described by \cite{Zhao2014} as an M dwarf. The amount of third light contributed to the system, affecting the accuracy of the transit parameter derivation, was determined over wavelength by \cite{MallonnStrass} and \cite{Nortmann2016}.

We monitored nearly a dozen eclipse events of the hot Jupiter HAT-P-32\,b with the 1.2 m telescope STELLA. Previous studies, \cite{Lendl2013} and \cite{Delrez2018}, showed that the small photometric signal can be revealed by meter-class telescopes when multiple observations are stacked. The new data of HAT-P-32\,b are employed to provide a follow-up study of the wavelength-dependence of the thermal flux by extending the previously probed wavelength range down to 0.9~$\mu$m. Furthermore, the same data are used to constrain the geometric albedo of the planet in the z' band.

The paper is organized as follows. Section~2 describes the observations and the photometric data reduction, Section~3 provides the data analysis and the result for the secondary eclipse depth. Section~4 presents the employed method to derive an upper limit on the geometric albedo and its result for the five targets of interest. A discussion of the results is supplied in Section~5, while Section~6 provides an outlook for the potential of future ground-based measurements. In Section~7, we summarize our work in our conclusions.

\section{Observations and data reduction}
We obtained 11 individual light curves of secondary eclipse events with the 1.2 m STELLA telescope \citep{Strassmeier2004} from 2012 to 2018. The employed instrument was the wide field imager WiFSIP, providing a field of view (FoV) of 22\,$'\times$ 22\,$'$ on a scale of 0.32$''$/pixel \citep{Granzer2010}. The detector is a single 4096$\times$4096 back-illuminated thinned CCD with 15 $\mu$m pixels. We slightly defocused the telescope to achieve more stable photometry and applied a read-out window to reduce the read out time.

\begin{figure*}
\includegraphics[width=\hsize]{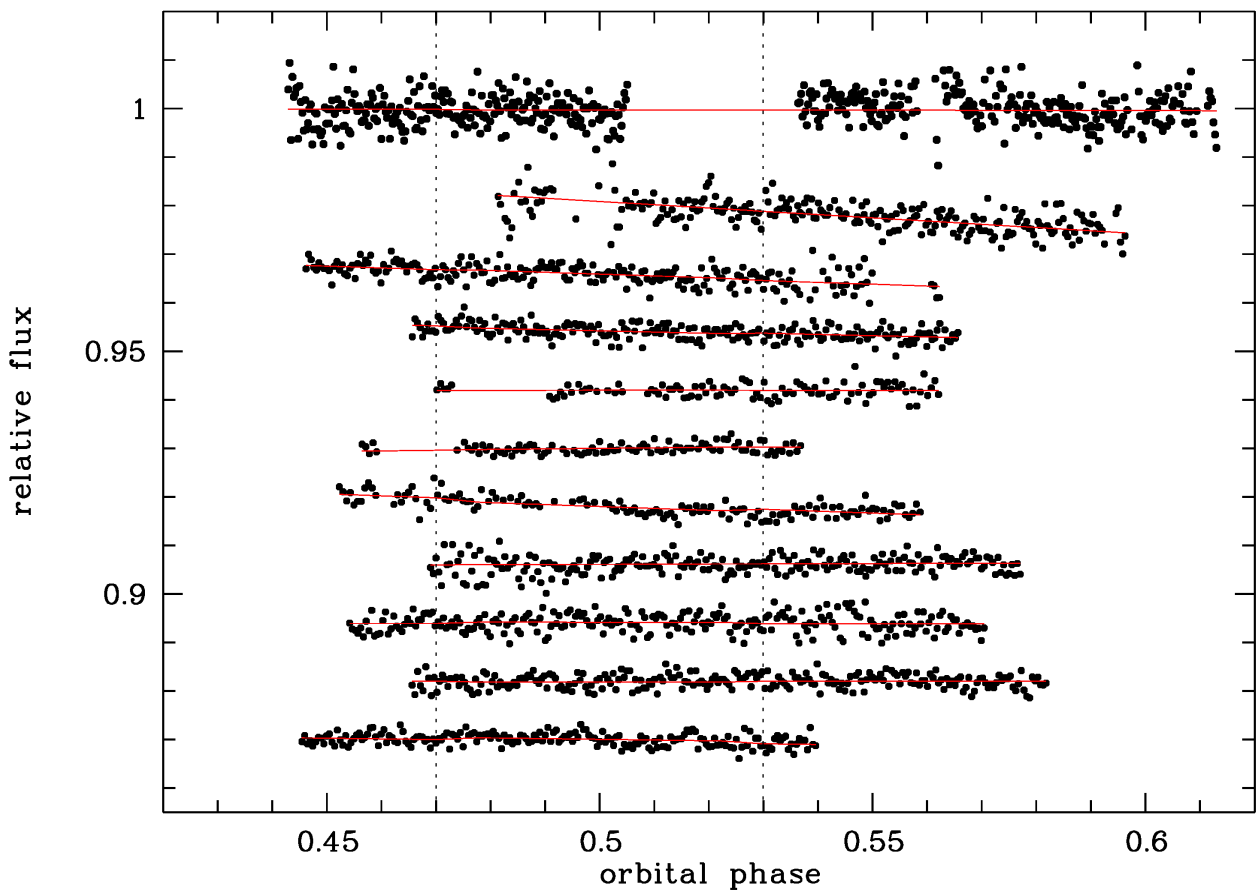}
\caption{Secondary eclipse light curves of HAT-P-32\,b, observed with STELLA/WiFSIP from 2012 to 2018. The red solid line shows the individual best-fit model including an eclipse and detrending. The dotted vertical lines denote beginning and end of the eclipse event. The sequence from top to bottom follows the order of Table \ref{tab_overview}.}
\label{plot_alllcs}
\end{figure*}

The data reduction followed the procedure of previous exoplanet light curve studies with STELLA \citep{Mallonn2015,Mallonn2016}. Bias and flat field correction was done by the STELLA pipeline. We used Sextractor \citep{Bertin96} for aperture photometry and extracted the flux in circular apertures of different sizes. ESO-Midas routines estimated the scatter of the light curves compared to a low-order polynomial over time and selected the aperture size that minimized the scatter. The same criterion of dispersion minimization was used to define the best selection of comparison stars per light curve. The 11 individual light curves are shown in Figure~\ref{plot_alllcs} and a summary of their properties is given in Table~\ref{tab_overview}.

\begin{table*}
\caption{Overview of secondary eclipse observations taken with the STELLA telescope in the Sloan z' filter. The columns provide the observing date, number of observed individual data points, exposure time, observing cadence, dispersion of the data points as root-mean-square (rms) of the observations after subtracting an eclipse model and a detrending function, the $\beta$ factor (see Section \ref{sec_ana}), and the airmass range of the observations.}
\label{tab_overview}
\begin{center}
\begin{tabular}{lrrrrrr}
\hline
\hline
\noalign{\smallskip}
Date & $N_{\mathrm{data}}$ &  $t_{\mathrm{exp}}$ (s) & Cadence (s) &   rms (mmag) &  $\beta$ &  Airmass \\
\hline
\noalign{\smallskip}
2012-10-26  & 494 & 20  &  48   &  3.25  &  1.69  &  1.05-2.43   \\
2012-11-23  & 235 & 50  &  78   &  2.46  &  1.19  &  1.05-2.23   \\
2013-01-05  & 216 & 60  &  88   &  1.78  &  1.16  &  1.05-2.47   \\
2013-01-18  & 210 & 60  &  88   &  1.38  &  1.00  &  1.05-2.07   \\
2015-01-06  & 99  & 90  &  123  &  1.37  &  1.00  &  1.13-2.47   \\
2015-02-03  & 94  & 90  &  123  &  0.95  &  1.28  &  1.13-2.32   \\
2016-10-15  & 137 & 90  &  123  &  1.38  &  1.68  &  1.05-1.74   \\
2017-10-19  & 214 & 60  &  93   &  1.78  &  1.19  &  1.05-1.76   \\
2017-11-03  & 230 & 60  &  93   &  1.83  &  1.08  &  1.05-1.32   \\
2017-11-16  & 229 & 60  &  93   &  1.39  &  1.26  &  1.05-1.58   \\
2018-01-13  & 188 & 60  &  93   &  1.20  &  1.40  &  1.05-1.87   \\
\hline                                                                                                     
\end{tabular}
\end{center}
\end{table*}

\section{Data analysis and result of eclipse depth}
\label{sec_ana}

To model our eclipse light curve we defined a simple trapezoid function similar to the procedure in \cite{vonEssen2015}. The model is defined by the four contact points of the eclipse, which result from the transit and ingress/egress duration and the transit midpoint. We fixed the former to the values of \cite{Zhao2014} and \cite{Nikolov2018} to allow for direct comparability, and for the latter we used the most recent ephemeris of \cite{Tregloan2018} under the valid assumption of zero eccentricity \citep{Zhao2014,Nikolov2018}. A summary of the adopted system parameters is given in Table~\ref{tab_param}.

\begin{table}
\caption{Input system parameters of HAT-P-32\,b used for the secondary eclipse model. From first to last row, it lists the orbital inclination $i$, orbital semimajor axis $a/R_{\star}$, planet-star radius ratio $R_p/R_{\star}$, eccentricity of the orbit $e$, argument of periastron $\omega$, zero point of the transit ephemeris $T_0$, and orbital period $P$. The parameters were adopted from \cite{Nikolov2018} and \cite{Tregloan2018} and were kept fixed in the analysis.}
\label{tab_param}
\begin{center}
\begin{tabular}{lr}
\hline
\hline
\noalign{\smallskip}
Parameter & Value \\
\hline
\noalign{\smallskip}
$i$ ($^{\circ}$) &  88.90  \\
$a/R_{\star}$    &  6.05  \\
$R_p/R_{\star}$  &  0.1508  \\
$e$              &  0.0   \\
$\omega$ ($^{\circ}$)         &  0.0   \\
$T_0$ $\mathrm{(BJD_{TDB})}$           &  2454420.447187  \\
$P$ (days)              &  2.1500080  \\
\hline                                                                                                     
\end{tabular}
\end{center}
\end{table}

Additionally to a potential eclipse dip, the ground-based light curves show smooth trends over time. We tested linear combinations of low-order detrending functions over external parameters like time, airmass, detector position, or full width half maximum of the point spread function and selected the best detrending model according to the Bayesian information criterion (BIC). As in previous transit light curve studies with STELLA/WiFSIP \citep{Mallonn2015,Mallonn2016,Alexoudi2018}, we found a low-order polynomial over time as the best option. Indeed, for all the 11 eclipse light curves presented in this work, the BIC was minimized by a linear function over time.

As a first step of the analysis, we applied a 4$\sigma$ outlier rejection to our data. In a next step, we ran an initial eclipse and detrending model fit and enlarged the photometric uncertainties by a common factor that the reduced $\chi^2$ value reached unity. Then, we calculated the so-called $\beta$ factor that takes into account correlated noise \citep{Gillon06,Winn08}. The photometric uncertainties were enlarged further by this factor. Details on this procedure are given in our previous transit light curve studies \citep{Mallonn2015,Mallonn2016}.

We fit the parameters of the eclipse plus detrending model by a Markov chain Monte Carlo (MCMC) approach that makes extensive use of PyAstronomy\footnote{https://github.com/sczesla/PyAstronomy}, developed by the PyA team at Hamburger Sternwarte. First, we fitted the 11 light curves individually. Free parameters were the depth of the secondary eclipse $d$ and the two coefficients $c_{0,1} $ of the linear detrending function over time. For each light curve, we ran a MCMC with 300.000 iterations, burned the first 100.000 steps, and used the mean and standard deviation of the parameter posterior distribution as best-fit values and its 1$\sigma$ uncertainties. We verified the convergence of the chains by dividing them into four parts after removal of the burn-in. The four mean values were consistent within their 1$\sigma$ errors for all light curves. 
Throughout our work, we allowed the eclipse depth to fluctuate around zero including physically meaningless negative values. Our intention was to avoid artificially pushing our result toward a detection by only allowing positive values, following numerous literature examples of this procedure \citep{Rowe2008,Evans2013,Bell2017}. Thus, for 4 of our 11 light curves, we obtain negative values of low significance for the eclipse depth (Table~\ref{tab_res_d}). 

\begin{figure}
\includegraphics[width=\hsize]{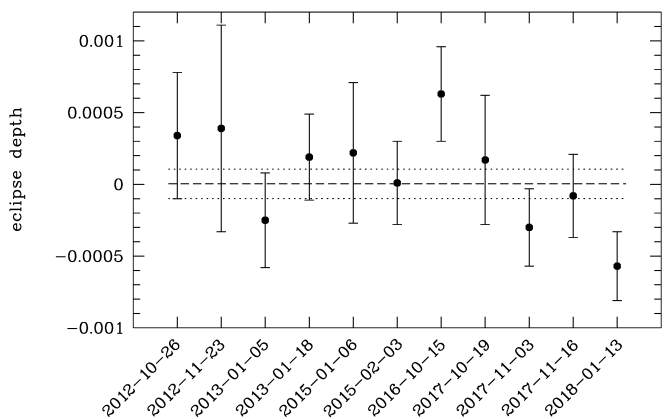}
\caption{Eclipse depth for the 11 individual observations. The horizontal dashed line depicts the best-fit eclipse depth for the joint fit of all data, the horizontal dotted lines indicate its 1$\sigma$ uncertainty.}
\label{plot_res}
\end{figure}

To check the robustness of the derived uncertainties on $d$, we calculated the reduced $\chi^2$ value under the assumption of no time variability of the eclipse depth. For 10 degrees of freedom, we achieve an $\chi^2_{\mathrm{red}}$ value of 1.27. While this value greater than unity hint at a mild effect of systematic errors, there is no indication of a significant underestimation of the uncertainties on $d$. Hence, the individual uncertainties are in rough agreement with the overall scatter of the results (see Figure~\ref{plot_res}). 

\begin{figure}
\includegraphics[width=\hsize]{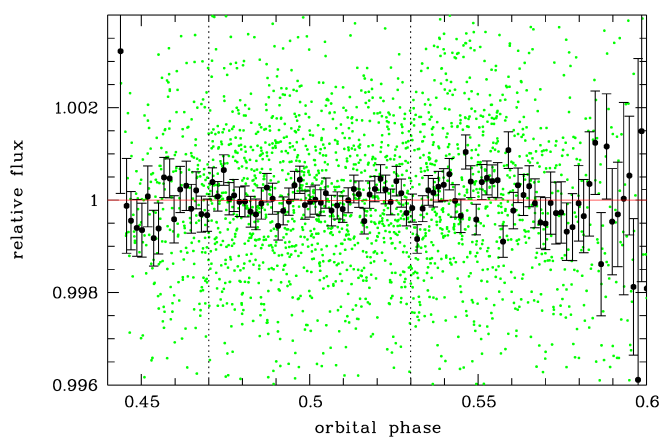}
\caption{Secondary eclipse light curve after orbital phase folding. The black data points indicate the values binned in 5~min intervals of the orbital phase. The green data points show the unbinned values of the 11 individual light curves after detrending. The best-fit model of the joint fit is given in red. The dotted vertical lines denote beginning and end of the eclipse event.  }
\label{plot_binlc}
\end{figure}

Subsequently, we fitted the data of all 11 light curves jointly with one common parameter $d$ for the eclipse depth and individual detrending coefficients. In total, the fit involved 23 free parameters. The result of \mbox{$d\,=\,-0.006\pm0.103$~ppt} for the joint fit is a null detection with a tight 97.5\% confidence upper limit of 0.196~ppt. We note that \cite{Zhao2014} measured the secondary eclipse events to happen within 2~min of the predicted zero-eccentricity timings. Thus, with a eclipse duration of about 186~min, our restriction to zero eccentricity has no influence. We are confident that we have not accidentally missed the eclipse intervals. For visualization purpose, we phase folded the data and binned them in 5~min steps; see Figure~\ref{plot_binlc}. The point-to-point scatter of this binned light curve between orbital phase 0.45 and 0.55 is 0.36~ppt. 

At this stage, our measurements still include the third light contribution of the M dwarf HAT-P-32\,B \citep{Adams2013} because its light is unresolved in our mildly defocused images. \cite{Zhao2014} reported a light contribution of the M dwarf in the z' band of 1.2\%, which was confirmed by \cite{MallonnStrass} and \cite{Nortmann2016}. According to Equation~3 in \cite{Zhao2014}, the secondary eclipse of the planet HAT-P-32\,b gets diluted by this quantity, thus the dilution corrected value is -0.006$\pm$0.104~ppt, with a 97.5\% confidence upper limit of 0.198~ppt.

We performed two additional tests on the robustness of the derived result. Firstly, we calculated an error-weighted mean of all the out-of-eclipse data and found this value to be in agreement with the weighted mean of the in-eclipse data at the $10^{-5}$ level. Secondly, we injected an eclipse signal into the original data of \mbox{$d\,=\,0.25$~ppt} and repeated all steps of the data analysis described
previously. We recovered this artificial signal to \mbox{$d\,=\,0.22\pm0.10$~ppt}, providing us confidence in the analysis procedure and the final null detection.

To determine an effective wavelength for our measurements, we multiplied the filter curve of the employed Sloan z' filter with the quantum efficiency curve of the CCD filter and a spectral energy distribution of a F8V star from the spectral flux library of \cite{Pickles1998}. The product of these three contributions rises steeply from zero to its maximum between 820 and 840~nm, from where it declines approximately linearly to zero at about 1000~nm. The effective wavelength is located at 890~nm.

\begin{table}
\caption{Eclipse depth $d$ for the 11 individual observations and the joint data set with 1$\sigma$ uncertainties including a correction for the dilution of HAT-P-32\,B.}
\label{tab_res_d}
\begin{center}
\begin{tabular}{lr}
\hline
\hline
\noalign{\smallskip}
Date & $d$ (ppt)  \\
\hline
\noalign{\smallskip}
2012-10-26  &  0.34 $\pm$ 0.44  \\
2012-11-23  &  0.39 $\pm$ 0.72  \\
2013-01-05  & -0.25 $\pm$ 0.33  \\
2013-01-18  &  0.19 $\pm$ 0.30  \\
2015-01-06  &  0.22 $\pm$ 0.49  \\
2015-02-03  &  0.01 $\pm$ 0.29  \\
2016-10-15  &  0.63 $\pm$ 0.33  \\
2017-10-19  &  0.17 $\pm$ 0.45  \\
2017-11-03  & -0.30 $\pm$ 0.27  \\
2017-11-16  & -0.08 $\pm$ 0.29  \\
2018-01-13  & -0.57 $\pm$ 0.24  \\
\hline
\noalign{\smallskip}
joint       & -0.01 $\pm$ 0.10  \\
\hline                                                                                                     
\end{tabular}
\end{center}
\end{table}

\section{Geometric albedo in the z' band}
\label{sec_albedo_H32}

Generally, the light we receive from an exoplanet is the combination of reflected light and thermally emitted light \citep{Alonso2018}. The depth of the secondary eclipse, which describes the relative contribution of the planetary flux to the stellar flux $F_p/F_S$, can be expressed as \citep{Seager2010} 
\begin{equation}
d_{\mathrm{refl}}\,=\,A_g\, \bigg(\frac{R_p}{a}\bigg)^2 \, , 
\label{equ_one}
\end{equation}
for the reflected light component, where $A_g$ is the geometrical albedo, $R_p$ is the planetary radius, and $a$ is the orbital semimajor axis. For the component of the thermally emitted planetary light, the eclipse depth is written as \citep{Seager2010}
\begin{equation}
d_{\mathrm{therm}}\,=\,\frac{B(\lambda,T_{d,p})}{B(\lambda, T_s)}\,\bigg(\frac{R_p}{R_s}\bigg)^2 \, , 
\label{equ_two}
\end{equation}
where $R_s$ is the radius of the host star, and $B(\lambda,T_{d,p})$ and $B(\lambda,T_s)$ are the blackbody emissions of the planetary day side and the star at the temperatures $T_{d,p}$ and $T_s$, respectively. Combining Equation~\ref{equ_one} and \ref{equ_two}, an observed value for the eclipse depth $d$ can be explained by the inversely related contributions of the reflected and thermal component. Figure~\ref{plot_A_T} shows the relationship between geometric albedo and brightness temperature for a given eclipse depth, similarly presented in, for example, \cite{Demory2011}, \cite{Gaidos2017}, and \cite{Keating2017}. Solving for $A_g$, we obtain
\begin{equation}
A_g\,=\,d\,\bigg( \frac{a}{R_p}\bigg)^2 \,-\,\frac{B(\lambda,T_{d,p})}{B(\lambda, T_s)}\,\bigg(\frac{a}{R_s}\bigg)^2 \, .
\label{equ_three}
\end{equation}
Therefore, under reasonable assumptions for the thermally emitted light of HAT-P-32\,b in the z' band, we can constrain the amount of reflected light and thus infer an upper limit on its geometric albedo at $\sim\,0.9$~$\mu$m.

\subsection{Albedo of HAT-P-32\,b}
\label{albedo_H32}
\begin{figure}
\includegraphics[width=\hsize]{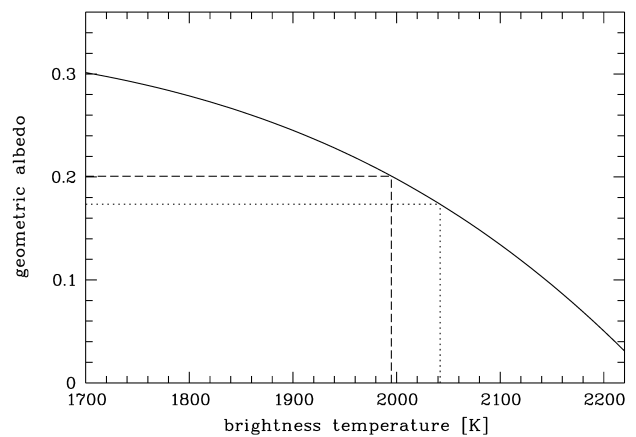}
\caption{Relationship between geometric albedo and brightness temperature. The curve shows the contour for the measured upper limit on the eclipse depth, $d\,<\,0.198$~ppt. The vertical dashed line corresponds to the planetary blackbody temperature determined by \cite{Nikolov2018} of $1995\pm17$~K, which is adopted in this work, and the vertical dotted line indicates the slightly larger value of \cite{Zhao2014}.}
\label{plot_A_T}
\end{figure}

To estimate an upper limit of the geometric albedo $A_g$ of \mbox{HAT-P-32\,b} in the z' band, we make the simplifying assumption that we can extrapolate the NIR emission spectrum measured by \cite{Nikolov2018} and \cite{Zhao2014} to the optical wavelengths as a blackbody spectrum. The two studies commonly described the NIR emission as best explained by a model with an isothermal P-T profile, and derived a blackbody temperature of $1995\,\pm\,17$~K, respectively $2042\,\pm\,50$~K. The more conservative upper limit on the geometric albedo is derived for the lower temperature owing to a minimized contribution of thermal emission; see Figure~\ref{plot_A_T}. Thus, we adopt a planetary blackbody radiation of 1995~K, and use our 97.5\% confidence upper limit on the eclipse depth of 0.198~ppt to infer an upper limit on the geometric albedo by Equation~\ref{equ_three}. The involved planetary parameters were adopted from \cite{Nikolov2018}. The result is $A_g\,<\,0.20$ (Fig.~\ref{plot_A_T}). 
The error budget is dominated by the uncertainty on $d$, thus the uncertainties of $R_p$, $R_s$, $a$, and $T_s$ can be neglected.

We provide arguments for the robustness of this upper limit, which relies on the trustworthiness of the emission extrapolation from the NIR to the z' band as a blackbody spectrum. In fact, the inferred value of $A_g$ would deviate for a thermal emission different from the blackbody emission. Because hot Jupiters have very different day and night sides, there is a temperature gradient in the day side of the planet approximately outward from the substellar point. Thus, different locations emit at different temperatures and the resulting emission do not follow a Planck function \citep{Schwartz2015}. In their Figure 6, \cite{Schwartz2015} showed that, indeed, the thermal flux is underestimated by the assumption of a Planck curve. In the case of a planet of 2000~K brightness temperature such as HAT-P-32\,b, the thermal component in the z' band gets underrated by about 10\%. However, as far as we underestimate the thermal component, our upper limit on $A_g$ remains robust because the true value might in fact be even lower. 

Generally, if the day side atmosphere is cloud-free at the probed pressure range and non-isothermal, we would expect to measure a wavelength-dependent brightness temperature due to spectral features. In this case, the z' band usually shows a larger eclipse depth and brightness temperature than the best-fit blackbody temperature \citep[][see also Fig.~\ref{plot_spec}]{Delrez2018}. As discussed above, in case of a larger z' band brightness temperature, our upper limit on $A_g$ would remain robust. Also, there is currently no observational indications for a non-isothermal profile of the day side of HAT-P-32\,b. For hot Jupiters in general, \cite{Schwartz2015} formed an aggregate broadband emission spectrum of the available measurements and found it to be flat and featureless, resembling a blackbody model. In case of a non-isothermal model, \cite{Cowan2011} predicted the z' band brightness temperature to differ within $\sim$\,100~K compared to the NIR broadbands. 

We note that neither \cite{Nikolov2018} nor \cite{Zhao2014} considered a component of reflected light in their NIR secondary eclipse measurements. Despite the predominance of the thermal component at wavelengths redward of $\sim$\,0.8~$\mu$m \citep{LopezMorales2007}, any amount of reflected light would, in principle, reduce the thermal component for a given eclipse depth and thus lower the inferred blackbody temperature. A decreased blackbody temperature would lead to a larger albedo value for the same eclipse depth (Figure~\ref{plot_A_T}), counteracting the robustness of our upper limit. To estimate the effect of a component of reflected light at the Spitzer eclipse events at 3.6 and 4.5~$\mu$m, we assumed a moderate geometric albedo of 0.2. The derived blackbody temperature decreased accordingly by only $\sim$\,30~K, which does not modify the extrapolated thermal component at optical wavelength significantly. In conclusion, we consider our upper limit on $A_g$ as robust.

\subsection{Albedo constraints for other hot Jupiters}
There are several literature reports of z' band secondary eclipse measurements for different planets \citep{Lopez2010,Lendl2013,Foehring2013,Delrez2016,Delrez2018}. These gas giants show equilibrium temperatures above 2500~K, hence they belong to the group of the so-called ultra-hot Jupiters. So far, these measurements have been solely interpreted according to their information content for the emission spectra. However, among the investigated targets, there are several objects with a potentially large reflection signal (see Section~\ref{sec_outlook}). Additionally, independent emission measurements at longer wavelengths exist, which can be used to extrapolate the thermal light component from the NIR to the z' band. In the following, we apply the same method as for HAT-P-32\,b in Section~\ref{albedo_H32} on z' band measurements of WASP-12\,b, WASP-19\,b, WASP-103\,b, and WASP-121\,b to constrain their geometric albedo.

\subsubsection{WASP-19\,b}
\label{sec_W19}

For WASP-19b, two independent studies measured the secondary eclipse depth in the z' band: \cite{Lendl2013} found $d\,=\,0.035 \pm 0.012$~ppt, while \cite{Burton2012} derived a value of $d\,=\,0.088 \pm 0.019$~ppt. The obtained two values are 2.4$\sigma$ in disagreement. One additional measurement was obtained by \cite{Zhou2013}, however this work achieved a result of lower precision. 

In order to infer the amount of reflected light and to derive a value for the geometric albedo, we extrapolate the thermal component from the NIR to the z' band by treating it as blackbody emission spectrum similar to our procedure for HAT-P-32\,b. We searched the literature for brightness temperature estimates of the day side of WASP-19b. Our review yielded three values: 2250~K \citep{Bean2013}, 2300~K \citep{Anderson2013}, and 2370~K \citep{Wong2016}. The most conservative upper limit on the geometric albedo is derived for the lower temperature owing to a minimized contribution of the thermal emission. Thus, we adopt \mbox{$T_{d,p}\,=\,2250$~K} and derive a 97.5\% confidence upper limit of \mbox{$A_g\,<\,0.21$} for the eclipse depth of \cite{Lendl2013}, and \mbox{$A_g\,<\,0.64$} for the \cite{Burton2012} eclipse depth. Planetary parameter values were taken from \cite{Lendl2013}. We note that the thermal emission amounts to $\sim$\,0.25~ppt at 0.9~$\mu$m assuming blackbody radiation of 2250~K, i.e., a dominating reflected light component would be required to achieve the eclipse depth $d$ of \cite{Burton2012}. Indeed, their $d$ value would result in a significant detection of the geometric albedo of \mbox{$A_g\,=\,0.4 \pm 0.12$}. However, we suggest caution in reviewing these findings because the result of \cite{Burton2012} is based on a single eclipse observation in contrast to the observations in this work and the z' band eclipse measurements of \cite{Lendl2013} and \cite{Delrez2018} (see Section~\ref{sec_W103}). Repeated observations are believed to not only increase the formal signal-to-noise ratio of the measurement, but also the accuracy and robustness of the result by averaging out systematic errors \citep{Bean2013,Lendl2013,Hansen2014,Mallonn2015}. In fact, for single Spitzer light curves, \cite{Hansen2014} suggested inflating the uncertainties derived by standard procedures. Therefore, we adopt in this work the upper limit obtained from the \cite{Lendl2013} measurement of $A_g\,<\,0.21$.

\subsubsection{WASP-12\,b}
\label{sec_W12}

For WASP-12\,b, there are two independent eclipse measurements in the z' band, which are not fully consistent. \cite{Lopez2010} found a secondary eclipse depth of $0.82\,\pm\,0.15$~ppt, while \cite{Foehring2013} derived a value of \mbox{$d\,=\,1.3 \pm 0.13$~ppt}. Corrected for third light, these values increase to $0.85\,\pm\,0.16$~ppt and $0.135\,\pm\,0.13$~ppt, respectively \citep{Stevenson2014}. Following \cite{Stevenson2014}, we adopt the value of \cite{Lopez2010} because of its broad consistency with modeling results from NIR to mid-infrared measurements, while the value of \cite{Foehring2013} appears to be largely inconsistent with any modeling attempts.

There is a general agreement about the derived best-fit blackbody temperatures for WASP-12\,b in the literature at NIR wavelengths \citep{Stevenson2014,Parmentier2018}, and we use the slightly lower value of 2894~K by \cite{Parmentier2018} to achieve a more conservative upper limit on the geometric albedo. The values of $R_S$, $R_p$, and $a$ were adopted from \cite{Collins2017}. 
We constrain the geometric albedo to \mbox{$A_g\,=\,0.17 \pm 0.11$} with an 97.5\% upper limit of 0.38 for the \cite{Lopez2010} depth. Similarly to the single-eclipse measurement of \cite{Burton2012} for \mbox{WASP-19\,b} (Section~\ref{sec_W19}), we suggest caution to both $d$ values of \cite{Lopez2010} and \cite{Foehring2013}. They are based on single observations, which appear to be less reliable than repeated observations. Thus, follow-up observations are desired.

\subsubsection{WASP-103\,b}
\label{sec_W103}
We reinterpret the z' band measurement of \cite{Delrez2018} of \mbox{$d\,=\,0.70 \pm 0.11$~ppt} for their information content on the geometric albedo. The estimated blackbody temperatures available in the literature are in broad agreement \citep{Kreidberg2018,Parmentier2018,Cartier2017}. Because of a conservative upper limit on the albedo, we choose the lowest of 2890~K derived by \cite{Cartier2017}. The values of the planetary parameters involved in Equation~\ref{equ_one} and \ref{equ_two} were taken from \cite{Delrez2018}. This results in a tight constraint of a 97.5\% confidence upper limit of $A_g\,<\,0.16$, the lowest value derived in this work (see Table~\ref{tab_limits_Ag}).

\subsubsection{WASP-121\,b}
\cite{Delrez2016} presented a z' band secondary eclipse measurement for WASP-121\,b of \mbox{$d\,=\,0.60 \pm 0.13$~ppt}. Two studies, \cite{Evans2017} and \cite{Parmentier2018}, fitted a blackbody model to Hubble Space Telescope WFC3 and Spitzer secondary eclipse data \citep{Garhart2019} and derived values in rough agreement to each other, $2700 \pm 10$~K and $2650 \pm 10$~K, respectively. For the same reasoning as in the previous chapters, we adopted the lower value and derived \mbox{$A_g\,=\,0.16 \pm 0.11$} with an 97.5\% upper limit of 0.37. Planetary parameter values were taken from \cite{Delrez2016}.

\begin{table}
\caption{Results for the 97.5\% confidence upper limit on the geometric albedo in the z' band derived in this work. The third column gives the literature reference for the z' band secondary eclipse measurement.}
\label{tab_limits_Ag}
\begin{center}
\begin{tabular}{lcr}
\hline
\hline
\noalign{\smallskip}
Planet & $A_g$ & Eclipse \\
       & upper limit & reference \\
\hline
\noalign{\smallskip}
HAT-P-32\,b & 0.20 & this work \\
WASP-12\,b & 0.38 & \cite{Lopez2010} \\
WASP-19\,b & 0.21 & \cite{Lendl2013} \\
WASP-103\,b & 0.16 & \cite{Delrez2018} \\
WASP-121\,b & 0.37 & \cite{Delrez2016} \\
\hline                                                                                                     
\end{tabular}
\end{center}
\end{table}

\section{Discussion}

\subsection{Emission spectrum of HAT-P-32\,b}
\begin{figure}
\includegraphics[width=\hsize]{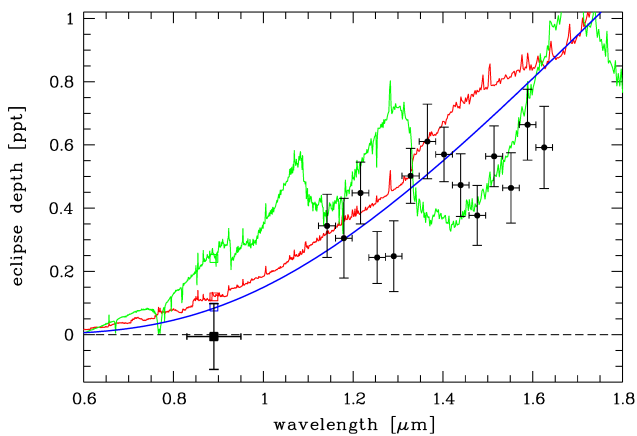}
\caption{Emission spectrum of HAT-P-32\,b. The z' band measurement of this work is shown as square symbol; the data points of \cite{Nikolov2018} are shown as circles. The horizontal lines denote the wavelength interval of each data point. Overplotted in red is an ATMO forward model with an inverted atmospheric P-T profile including TiO/VO absorption, in green a model with a non-inverted atmosphere without TiO/VO (both reproduced from \cite{Nikolov2018}), and in blue a blackbody spectrum for star and planet with a planetary temperature of 1995~K. The model values binned to the z' band wavelength interval are shown with open squares.}
\label{plot_spec}
\end{figure}

In a previous work on the day side emission spectrum of \mbox{HAT-P-32\,b}, \cite{Nikolov2018} combined \textit{HST} WFC3 secondary eclipse data from 1.12 $\mu$m to 1.64~$\mu$m with H and K$_\mathrm{S}$ band and \textit{Spitzer} data at 3.6 $\mu$m and 4.5~$\mu$m of \cite{Zhao2014}. They compared the data to forward models computed with the \textit{ATMO} code \citep{Goyal2018}. Among others, one model was calculated using a P-T profile with temperature decreasing with altitude (non-inverted) and with TiO/VO removed. A second model exhibited a P-T profile of temperature increasing with altitude (inverted), containing TiO/VO. Our z' band data point confirms their result that the model with the non-inverted \mbox{P-T} profile lacking TiO/VO shows a poor match to the data (Figure~\ref{plot_spec}). This model is disfavored by our measurement by 2.5$\sigma$. \cite{Nikolov2018} and \cite{Zhao2014} found the best-fit planetary model to be one of blackbody radiation, i.e., a model with an isothermal P-T profile, which is not uncommon for hot Jupiter emission spectra \citep[e.g.,][]{Hansen2014}. Our z' band measurement of this work is in 1$\sigma$ agreement to a blackbody spectrum of the best-fit temperature of 1995~K derived by \cite{Nikolov2018} using all data from 1.12 $\mu$m to 4.5~$\mu$m.  

\subsection{Implication for clouds in atmosphere of HAT-P-32\,b}

The isothermal P-T profile in the planet day side derived by \cite{Nikolov2018} might potentially be explained by clouds formed by silicate condensates. \cite{Mallonn2017} found indications for this type of condensates at the terminator region of the same planet. Early theoretical calculations predicted that such a cloud layer, if present at the planets dayside, can cause a geometric albedo of $\sim$~0.4 over most of the optical to very NIR wavelength region \citep{Sudarsky2000}. This predicted albedo value is ruled out by our estimation for HAT-P-32\,b. Interestingly, for the planet Kepler-7b, similar to HAT-P-32\,b in terms of surface gravity and equilibrium temperature \citep{Latham2010}, an albedo $A_g\,\sim\,0.35$ was derived in the Kepler band and interpreted as reflection on a silicate cloud deck \citep{Demory2013}. HAT-P-32\,b does not possess a similarly large albedo at 0.89~$\mu$m. However, \cite{Sudarsky2000} used a condensate grain size distribution peaking at 5~$\mu$m. \cite{GarciaMunoz2015} modeled the clouds of Kepler-7b with smaller condensate particles and found a wavelength-dependence due to Mie-scattering. Hence, while our albedo upper limit can rule out large-particle clouds, it cannot rule out small-particle clouds since the geometric albedo might be as small as our derived limit. 

In a comparison between blackbody brightness temperature and theoretical equilibrium temperature, \cite{Nikolov2018} ruled out a low-Bond albedo low-recirculation scenario for the day side atmosphere. A low albedo, as suggested by our measurement, involves an energy recirculation efficiency $\varepsilon$, per definition ranging between zero and unity, larger than 0.4, which is not unusual for hot Jupiters of this temperature range \citep{Nikolov2018}. There is a tendency toward a lower recirculation efficiency as the stellar irradiation increases \citep{Cowan2011,PerezBecker2013}; however, HAT-P-32\,b does not belong to the planets that are exposed to large stellar irradiation, thus this planet does not counteract the trend.

\subsection{Sample of five z' band geometric albedo limits}
While for HAT-P-32\,b, WASP-103\,b, and WASP-121\,b this work constrains their geometric albedo for the first time, there are published optical albedo measurements for WASP-12\,b and WASP-19\,b. \cite{Bell2017} derived a very tight upper limit for WASP-12\,b of $A_g<0.064$ at $\lambda < 570$~nm with HST/STIS observations. Our upper limit is less precise but in general agreement with a low value. The Bond albedo of WASP-12\,b as estimated by NIR brightness temperatures might be larger than this, 0.3 to 0.4 \citep{Schwartz2015}. However, \cite{Schwartz2017} presented an alternative option for WASP-12\,b with a lower Bond albedo of 0.06.

\cite{Abe2013} measured an optical eclipse light curve of WASP-19\,b with the \textit{ASTEP400} telescope in Antarctica, deriving a geometric albedo of $0.27\,\pm\,0.13$ in a wavelength range of 575 to 760~nm without a correction for the thermally emitted flux by the planet. If we apply the correction assuming blackbody radiation of 2250~K, we obtain a corrected value of $A_g\,=\,0.20 \pm 0.13$. This is in agreement to the upper limit derived here of $A_g < 0.21$. From NIR observations, \cite{Wong2016} suggested a higher Bond albedo of $A_B = 0.38 \pm 0.06$. 

All geometric albedo limits derived in this work point toward low values, indicative of absorption acting at the wavelength region of about 0.9~$\,\mu$m probed here. We find no hints for reflective clouds in the day side atmospheres. For the four ultra-hot Jupiters, this is in line with theoretical observations predicting these atmospheres to be too hot to form clouds \citep{Parmentier2018,Wakeford2017}. Our observations do not shed light on the apparent discrepancy between the often moderate Bond albedos for hot Jupiters \citep[$A_B \approx 0.3-0.4$][]{Schwartz2015,Schwartz2017} and their low to very low geometric albedo values ($A_g \lesssim 0.2$) at optical wavelengths. \cite{Schwartz2015} suggested reflective clouds with additional optical absorbers as explanation. In case of TiO as absorber, their opacity is lower in the z' band than at optical wavelengths \citep{Cowan2011}. However, we do not find indications for larger z' band albedo values compared to optical values. The pressure-broadened wings of potassium partly overlap with the Sloan z' filter. Thus, the alkali absorption opacity might also contribute to a low value of the geometric albedo \citep{Burrows2008}. For the hottest of our planets, other optical opacities, for example bound-free absorption by H$^-$ \citep{Parmentier2018,Arcangeli2018}, are expected to play a role as well.

\section{Outlook for ground-based optical albedo measurements}
\label{sec_outlook}
The precision achieved in this work of 0.1~ppt is similar to comparable work on ground-based z' band eclipse photometry \citep{Lendl2013,Delrez2016,Delrez2018}. We estimate the number of known exoplanets for which this precision would be sufficient to detect a geometric albedo with 3$\sigma$ confidence, if we assume $A_g$ to be 0.4 as measured for HD189733\,b \citep{Evans2013}. We employed the catalog data of the TEPCAT\footnote{http://www.astro.keele.ac.uk/jkt/tepcat/tepcat.html} \citep{Southworth2011} to calculate the expected secondary eclipse depth of all known transiting exoplanets. We restricted the catalog to systems with a V magnitude brighter than 14. Our search resulted in 12 planets for which the eclipse depth corresponding to $A_g\,=\,0.4$ could be detected with about 3$\sigma$ if the same precision is reached as in our work (Table~\ref{tab_eclranking}). 

In the z' band, we achieved the high precision with a combined analysis of 11 photometric time series of the moderately bright star HAT-P-32. However, for optical bands, the quantum efficiency of most CCD detectors is higher than in the z' band, i.e., the same precision might be reached with a lower number of observations. Also, the usage of middle-sized telescopes, compared to the small-sized telescope used in the present study, allows for a similar precision with fewer observations. We emphasize again that, while more expensive in terms of telescope time, a large number of observations offers the advantage of a more robust result (see Section~\ref{sec_W19}). 

In principle, ground-based observations are able to reveal a wavelength dependence of the optical geometric albedo if the data are collected in multiple broadband filters. An early example of this idea was presented by \cite{Winn08}. Thus, ground-based optical eclipse observations can be complementary to space-based spectro-photometry \citep{Evans2013} and the proposed method of ground-based high-resolution spectroscopy \citep{Martins2018}.

\begin{table}
\caption{Transiting planets listed by their calculated eclipse depth for an assumed geometric albedo of 0.4. For these planets, a 3$\sigma$ detection is possible assuming a similar precision of 0.1~ppt as achieved in this work for HAT-P-32\,b.}
\label{tab_eclranking}
\begin{center}
\begin{tabular}{lr}
\hline
\hline
\noalign{\smallskip}
Planet & $d$ \\
 & (calculated, ppt) \\
\hline
\noalign{\smallskip}
WASP-19\,b    &  0.65  \\
WASP-103\,b   &  0.60  \\
WASP-12\,b    &  0.57  \\
HATS-18\,b    &  0.50  \\
WASP-121\,b   &  0.47  \\
KELT-16\,b    &  0.41  \\
WASP-43\,b    &  0.40  \\
WASP-33\,b    &  0.36  \\
CoRoT-1\,b    &  0.32  \\
WASP-4\,b     &  0.30  \\
WASP-18\,b    &  0.30  \\
HATS-24\,b    &  0.30  \\
\hline                                                                                                     
\end{tabular}
\end{center}
\end{table}

\section{Conclusions}
\label{sec5}

In this work, we presented the analysis of 11 secondary eclipse light curves of the hot Jupiter HAT-P-32\,b. A joint model fit to all light curves, including a simultaneous, individual detrending, resulted in an eclipse depth of \mbox{$d\,=\,-0.006\pm0.104$~ppt} with a 97.5\% confidence upper limit of 0.198~ppt. This null detection is in 1$\sigma$ agreement to the isothermal planetary emission model favored by the previous work of \cite{Nikolov2018} and disfavors the planetary emission model with a non-inverted P-T profile by 2.5$\sigma$ confidence.

Assuming the isothermal planetary emission model, resembling blackbody radiation, we can convert the upper limit on the eclipse depth for HAT-P-32\,b to an upper limit on the geometric albedo in the z' band. We achieve a value of \mbox{$A_g\,<\,0.20$}. Thus, a reflective silicate cloud, which might be an explanation for the isothermal emission spectrum, that has a grain size of about 5~$\mu$m is unlikely as it is predicted to cause an albedo of about 0.4 \citep{Sudarsky2000}. However, smaller grain sizes would certainly cause a wavelength-dependent albedo with lower values toward the z' band, thus this possibility might agree with our measurement.

We apply the same method of deriving an upper limit on the geometric albedo to z' band eclipse measurements published in the literature. This exercise comprises the targets WASP-12\,b, WASP-19\,b, WASP-103\,b, and WASP-121\,b, which all belong to the class of ultra-hot Jupiters. For two of the targets, WASP-103\,b, and WASP-121\,b, these are the first published albedo measurements. The derived 97.5\% confidence upper limits range from the tight value of $A_g\,<\,0.16$ for WASP-103\,b to the more loose value of $A_g\,<\,0.38$ for WASP-12\,b. In general, all z' band albedo constraints for the five hot to ultra-hot Jupiters investigated in this work  point toward low albedo values. Thus, our upper limits agree with the generally low values of optical geometric albedos derived at bluer wavelengths. We find no indications for clouds on the day side hemispheres of the planets, which otherwise might be revealed by their reflective properties. Our z' band albedo constraints confirm an apparent offset between Bond albedos deduced from NIR thermal phase curves \mbox{($A_B \approx 0.3-0.4$)} and geometric albedos obtained at shorter wavelengths \citep[$A_g \lesssim 0.2$,][]{Schwartz2015}. We suggest obtaining albedo measurements resolved in wavelength to gain further information on reflective clouds and estimate that ground-based broadband observations, even with small telescopes as done in this work, are a feasible option for a dozen of targets.

\begin{acknowledgements}
We thank Nikolay Nikolov and Jayesh Goyal for sharing their ATMO emission spectroscopy models of HAT-P-32\,b with us. We thank the anonymous referee for an insightful review that improved the manuscript. This work made use of PyAstronomy, the SIMBAD data base, and VizieR catalog access tool, operated at CDS, Strasbourg, France, and the NASA Astrophysics Data System (ADS).
\end{acknowledgements}

%
\bibliographystyle{aa} 
\bibliography{H32ecl_bib.bib} 
%

 

\end{document}